\documentclass[aps,twocolumn,superscriptaddress]{revtex4}
\usepackage{graphicx}

\newcommand{\bra}[1] {\left\langle #1 \right|}
\newcommand{\ket}[1] {\left| #1 \right\rangle}

\begin{document}

\title{Control-free control: manipulating a quantum system using
only a limited set of measurements}

\author{S. Ashhab}
\affiliation{Advanced Science Institute, The Institute of Physical
and Chemical Research (RIKEN), Wako-shi, Saitama 351-0198, Japan}
\affiliation{Physics Department, The University of Michigan, Ann
Arbor, Michigan 48109-1040, USA}

\author{Franco Nori}
\affiliation{Advanced Science Institute, The Institute of Physical
and Chemical Research (RIKEN), Wako-shi, Saitama 351-0198, Japan}
\affiliation{Physics Department, The University of Michigan, Ann
Arbor, Michigan 48109-1040, USA}

\date{\today}


\begin{abstract}
We present and discuss different protocols for preparing an
arbitrary quantum state of a qubit using only a restricted set of
measurements, with no unitary operations at all. We show that an
arbitrary state can indeed be prepared, provided that the
available measurements satisfy certain requirements. Our results
shed light on the role that measurement-induced back-action plays
in quantum feedback control and the extent to which this
back-action can be exploited in quantum-control protocols.
\end{abstract}

\maketitle

\section{Introduction}

Control techniques are applied in a wide variety of practical
applications, e.g. when a device is to be maintained in a certain
state in spite of the environmental fluctuations that would
normally push it away from that state \cite{ControlSystemBook}.
The basic structure of a closed-loop control system contains two
steps: measurement and feedback control. In the first step, i.e.
the measurement, information is acquired about the state of the
system and how far it is from the desired target state. In the
second step, a ``control'' is applied to the system, i.e. a signal
or force is applied, in order to change the state of the system
and guide it towards the target state.

In classical mechanics, the measurement process only extracts
information about the state of the system, but (under ideal
circumstances) it does not change that state. In quantum
mechanics, this picture breaks down: the measurement itself will
change the state of the system no matter how ideal it is (Note
that for any measurement outcome there are quantum states that are
not affected by the measurement; this absence of back-action,
however, cannot be true for a general state). One therefore needs
to treat quantum-control problems using a different frame of mind
from that used when dealing with classical-control problems
\cite{WisemanBook,Brif,Peirce}.

One possibility for dealing with the unavoidable back-action of
the measurement is to calculate the effected change and design the
control signal accordingly \cite{Belavkin,Wiseman}. Another
possibility, which might be conceptually more radical, is to use
the change caused by the measurement as the sole means for
manipulating the state of the system. In this case, closed-loop
feedback control involves only the measurement step; the
``control'' is no longer needed. Indeed there have been some
studies on this possibility in the past few years
\cite{Roa,Pechen,Jacobs}.

In previous work \cite{Roa,Pechen,Jacobs} it was assumed that
measurements in any basis are allowed. Here we consider the case
where only a restricted set of measurements is implementable, and
we analyze various questions related to whether such a limited set
of operations is sufficient for the preparation of an arbitrary
target state. We also consider the question of the time required
to reach the desired target state.

\section{State preparation using the full set of possible measurements}

In Refs.~\cite{Roa,Pechen} the authors considered the problem of
preparing an arbitrary target state from an arbitrary initial
state without imposing any constraints on the measurements that
can be performed on the system. There it was demonstrated that an
arbitrary target state can indeed be prepared using only
measurements. One can understand this situation as follows. For
any target state $\ket{\psi_T}$, one can construct (at least
formally) a projective measurement where $\ket{\psi_T}$ is one of
the possible outcomes. If now the above-described measurement is
performed on a system in any initial state, there is a possibility
that the outcome will correspond to the state $\ket{\psi_T}$, and
one would have succeeded in preparing the target state. This
procedure is obviously probabilistic; it is possible that one
might obtain a different outcome in the measurement. Turning this
probabilistic protocol into a deterministic one is
straightforward: every time the measurement fails to produce the
desired outcome, a perturbation, which can be thought of as a
kick, can be applied to the system and the measurement is
repeated. Unless the perturbation does not create any population
in the state $\ket{\psi_T}$, e.g.~for symmetry reasons, this
procedure is deterministic; if one keeps trying, one will
eventually obtain the target state \cite{Hida}.

In an increasingly large Hilbert space, it becomes more and more
unlikely to obtain the desired measurement outcome, which in turn
leads to longer and longer average state-preparation times. This
problem can be alleviated by making a better choice of
measurements than the one described above. One can guide the
quantum state of the system from the initial state to the target
state using a sequence of projective measurements where one of the
possible outcomes gradually changes from the initial state
$\ket{\psi_I}$ to the target state $\ket{\psi_T}$. For example,
one could design a sequence of $N$ projective measurements where
each measurement (labelled by the index $i$) has one outcome that
corresponds to the projection $\ket{\psi_i}\bra{\psi_i}$, where
\begin{equation}
\ket{\psi_i} = \cos\left(\frac{\pi i}{2N}\right) \ket{\psi_I} +
\sin\left(\frac{\pi i}{2N}\right) \ket{\psi_T}.
\end{equation}
For best performance, we assume that the other outcomes of the
measurement are orthogonal to $\ket{\psi_i}$. The probability that
the system will follow the state $\ket{\psi_i}$ in all the
measurement steps is given by $\cos^{2N}\left[\pi/(2N)\right]$,
which approaches unity in the limit $N\rightarrow\infty$,
regardless of the size of the Hilbert space. This procedure was
analyzed in Refs.~\cite{Roa,Pechen,vonNeumann}.

In Ref.~\cite{Jacobs} the author considered the case of continuous
measurement, where one essentially performs weak measurements
rather than projective measurements. There it was demonstrated
that by continuously adjusting the measurement settings, an
arbitrary target state can be prepared. The choice of the
measurement is done as follows: based on the instantaneous quantum
state and the desired target state, one chooses a measurement
basis that gives a high probability for the quantum state to
evolve towards the target state.

\section{State preparation using a restricted set of measurements}

In the previous section, we reviewed a number of ideas that can be
used for the preparation of an arbitrary target state using the
set of {\it all} possible measurements as available resources. In
the following, we consider the possibility of preparing an
arbitrary target state using a {\it small number} of available
measurements. Indeed, in realistic situations there typically are
constraints on the measurements that can be performed.

\subsection{Measurements of spin along three orthogonal axes}

We consider a two-level system, i.e.~a qubit, and we start by
considering measurements of this (pseudo-)spin along three
orthogonal axes, i.e.~the observables $\sigma_{x,y,z}$. If the
measurements were projective, the $\sigma_x$ measurement would
result in one of the states $\ket{\sigma_x=1}$ and
$\ket{\sigma_x=-1}$, depending on the outcome of the measurement,
and similarly for the two other observables. As a result, only six
target states can be prepared: $\ket{\sigma_x=1}$,
$\ket{\sigma_x=-1}$, $\ket{\sigma_y=1}$, $\ket{\sigma_y=-1}$,
$\ket{\sigma_z=1}$ and $\ket{\sigma_z=-1}$ (Below, we shall use
the notation $\ket{\uparrow}=\ket{\sigma_z=1}$ and
$\ket{\downarrow}=\ket{\sigma_z=-1}$).

\begin{figure}[h]
\includegraphics[width=5.0cm]{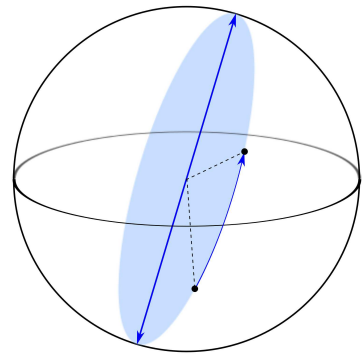}
\caption{(Color online) Schematic diagram showing the
transformation of the quantum state of a qubit as a result of a
non-projective measurement. The double arrow represents the
measurement axis. The tail of the single-headed arrow represents
the pre-measurement state, and the head represents the
post-measurement state. The quantum state ``moves'' in the plane
defined by the pre-measurement state and the measurement axis. The
distance that the quantum state moves depends on the different
parameters in the problem, including in particular the measurement
strength $\epsilon$.}
\label{Fig:SchematicSingleMeasurement}
\end{figure}

The situation changes drastically when, instead of strong
projective measurements, we consider weak, non-projective
measurements. An example of a weak measurement could be the
following: A measurement of $\sigma_z$ produces the outcome $+1$
or $-1$, upon which the quantum state of the system is transformed
(or, in some sense, partially projected) according to the formula
\begin{equation}
\rho_{\rm after} = \frac{\hat{M}_{\pm} \rho_{\rm before}
\hat{M}_{\pm}}{{\rm Tr} \{ \hat{M}_{\pm} \rho_{\rm before}
\hat{M}_{\pm} \}},
\end{equation}
where the measurement operators $\hat{M}_{\pm}$ are given by
\begin{eqnarray}
\hat{M}_+ & = & \sqrt{\frac{1+\epsilon}{2}}
\ket{\uparrow}\bra{\uparrow} + \sqrt{\frac{1-\epsilon}{2}}
\ket{\downarrow}\bra{\downarrow} \nonumber
\\
\hat{M}_- & = & \sqrt{\frac{1-\epsilon}{2}}
\ket{\uparrow}\bra{\uparrow} + \sqrt{\frac{1+\epsilon}{2}}
\ket{\downarrow}\bra{\downarrow},
\end{eqnarray}
and the probability of observing the two different outcomes are
given by
\begin{equation}
{\rm Prob}_{\pm} = {\rm Tr} \{ \hat{M}_{\pm} \rho_{\rm before}
\hat{M}_{\pm} \}.
\end{equation}
The parameter $\epsilon$ quantifies the strength of the
measurement: for a weak measurement $\epsilon$ is small, whereas
for a projective measurement $\epsilon=1$ (The parameter
$\epsilon$ can also be understood as the measurement fidelity
\cite{Ashhab}). We shall assume similar measurement properties for
the two other observables. A weak $\sigma_z$ measurement does not
project the system onto one of the two states $\ket{\uparrow}$ and
$\ket{\downarrow}$, but rather slightly modifies the quantum state
such that it experiences a small shift from the pre-measurement
state in the direction of one of the states $\ket{\uparrow}$ and
$\ket{\downarrow}$. For example, if one starts with the state
\begin{eqnarray}
\ket{\psi}_{\rm before} & = & \frac{1}{\sqrt{2}} \left(
\ket{\uparrow} + \ket{\downarrow} \right) \nonumber \\
& = & 0.707 \ket{\uparrow} + 0.707 \ket{\downarrow},
\end{eqnarray}
and one obtains the outcome $+1$ in a $\sigma_z$ measurement of
strength $\epsilon=0.01$, the post-measurement state will be
(approximately)
\begin{equation}
\ket{\psi}_{\rm after} = 0.711 \ket{\uparrow} + 0.703
\ket{\downarrow}.
\end{equation}
The transformation of the quantum state of a qubit following a
non-projective measurement is illustrated in
Fig.~\ref{Fig:SchematicSingleMeasurement}.

One can now observe that after a weak measurement of a given
observable, the post-measurement state lies in the plane defined
by the pre-measurement state and the measurement axis. Using this
observation, one can devise a protocol for preparing an arbitrary
target state using the three measurements mentioned above. One
possibility is the following: One performs a sequence of
$\sigma_x$ and $\sigma_y$ measurements until the quantum state of
the system lies in the plane defined by the $z$-axis and the
target state. Once that goal is achieved, one performs a
$\sigma_z$ measurement (or measurements), such that one either
obtains the target state or one concludes that the state has gone
too far in the opposite direction and is unlikely to come back
(clearly the above is a subjective explanation; however, it can be
made quantitative straightforwardly, as we shall do shortly). As a
general rule, one could say that success and failure in this
(i.e.~second) step occur with probability 50\% each. In the case
of failure, one goes back to the $\sigma_x$ and $\sigma_y$
measurements. This time one requires again that the state lie in
the plane defined by the $z$-axis and the target state, but one
also requires that the state approach the $x$-$y$ plane to within
a certain tolerance (this new condition is designed to bring the
state back to roughly the middle between the states
$\ket{\uparrow}$ and $\ket{\downarrow}$, such that the next
attempt at a $\sigma_z$ measurement will have a 50\% success
probability).

\begin{figure}[h]
\includegraphics[width=7.0cm]{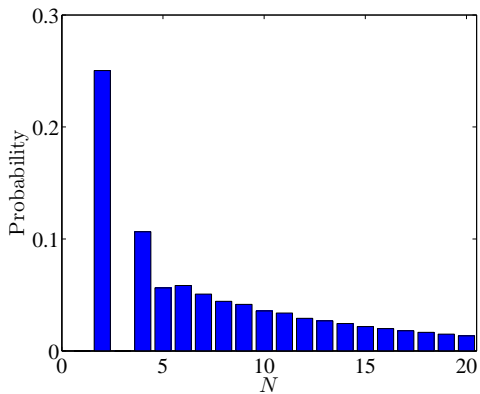}
\caption{(Color online) The probability of obtaining the target
state $\ket{\psi_T} = \cos\frac{\pi}{8} e^{-i\pi/8} \ket{\uparrow}
+ \sin\frac{\pi}{8} e^{i\pi/8} \ket{\downarrow}$ in $N$
measurement steps starting from the initial state $\ket{\psi_I} =
(e^{-i\pi/4} \ket{\uparrow} + e^{i\pi/4} \ket{\downarrow}
)/\sqrt{2}$ and following the procedure explained in Sec.~III.A.
For this particular choice of states, the average number of
measurement steps is 10.5 and the probability of successful state
preparation in under twenty steps is 0.86.}
\label{Fig:HittingTime}
\end{figure}

We now estimate the time required in order to prepare an arbitrary
quantum state using the above protocol. For this purpose we assume
that the measurement fidelity $F$ in a single measurement is a
tunable parameter (If one is constrained to use measurements of a
small fidelity $\epsilon$, any larger fidelity $F$ can be obtained
by repeating the low-fidelity measurement $n$ times with $n$ given
by
\begin{equation}
F={\rm erf} [\epsilon\sqrt{n\pi}/2] \approx 1-\exp\{-n\epsilon\},
\end{equation}
and ${\rm erf}$ stands for the error function \cite{Ashhab}). The
target state can be expressed as
\begin{equation}
\ket{\psi_T} = \cos\frac{\theta_T}{2} e^{-i\phi_T/2}
\ket{\uparrow} + \sin\frac{\theta_T}{2} e^{i\phi_T/2}
\ket{\downarrow},
\end{equation}
up to an irrelevant overall phase. For definiteness we assume that
the initial state points along the $y$-axis, i.e.~the initial
state is an eigenstate of $\hat{\sigma}_y$. As explained above,
before performing any $\sigma_z$ measurements, one first needs to
prepare the state
\begin{equation}
\ket{\psi} = \frac{1}{\sqrt{2}} \left( e^{-i\phi_T/2}
\ket{\uparrow} + e^{i\phi_T/2} \ket{\downarrow} \right).
\end{equation}
A measurement of $\sigma_x$ with fidelity $F=|\cos\phi_T|$ results
in the desired state with probability 0.5. In the case of failure,
a strong (and ideally projective) $\sigma_y$ measurement is
performed, followed by a new attempt using the $\sigma_x$
measurement (One should note here that the strong $\sigma_y$
measurement could result in a state with a value of $\phi$ that
differs from $\phi_T$ by more than $\pi/2$. In this case, it is
impossible for any $\sigma_x$ measurement to produce the desired
value of $\phi$. One could deal with this case by performing a
strong $\sigma_x$ measurement, followed by a $\sigma_y$
measurement with a properly calibrated fidelity). One can
therefore estimate that obtaining a state with the desired value
of the phase $\phi_T$ will, on average, require just a few
measurement steps. We now have a state that lies in the $x$-$y$
plane with the correct phase between the states $\ket{\uparrow}$
and $\ket{\downarrow}$. At this point, a $\sigma_z$ measurement
with fidelity $F=|\cos\theta_T|$ results in the target state with
probability 0.5. In other words, every time one succeeds in
obtaining the desired value of $\phi_T$, one has a 50\% chance of
obtaining $\ket{\psi_T}$ in the ensuing $\sigma_z$ measurement. In
the case where the $\sigma_z$ measurement fails to produce the
target state, one performs a strong $\sigma_y$ measurement and
goes back to the first step of the procedure explained above.
Since each attempt at preparing the target state involves a few
measurements (roughly three to five), and each attempt has a
success probability of 0.5, one can conclude that one has a high
probability of preparing the target state $\ket{\psi_T}$ in under
twenty measurement steps. In Fig.~\ref{Fig:HittingTime} we plot
the probability histogram for the number of measurement steps
required to prepare the target state $\ket{\psi_T} =
\cos\frac{\pi}{8} e^{-i\pi/8} \ket{\uparrow} + \sin\frac{\pi}{8}
e^{i\pi/8} \ket{\downarrow}$ starting from the initial state
$\ket{\psi_I} = (e^{-i\pi/4} \ket{\uparrow} + e^{i\pi/4}
\ket{\downarrow} )/\sqrt{2}$.

\subsection{Single measurement setting: symmetric, informationally complete, positive operator
valued measure}

Another interesting setup to consider is that of a qubit measured
using a symmetric, informationally complete, positive operator
valued measure (SIC-POVM). The reason why informationally complete
POVMs are interesting in this context is that they can (partially)
project the state of the system towards one of four different
directions that span the entire Bloch sphere, giving the quantum
state the possibility of moving in all directions about the Bloch
sphere.

The SIC-POVM on a single qubit is a measurement with four possible
outcomes where the four different outcomes correspond to quantum
states that form a tetrahedon on the Bloch sphere. One
representative choice of a SIC-POVM, which is equivalent to any
other SIC-POVM of the same strength up a rotation, is the one with
the following measurement operators:
\begin{equation}
\hat{M}_i = \frac{\sqrt{1+\epsilon}}{2} \ket{\psi_i}\bra{\psi_i} +
\frac{\sqrt{1-\epsilon}}{2}
\ket{\overline{\psi}_i}\bra{\overline{\psi}_i}
\end{equation}
with
\begin{eqnarray}
\ket{\psi_1} & = & \ket{\uparrow} \nonumber
\\
\ket{\psi_2} & = & \sqrt{\frac{1}{3}} \ket{\uparrow} +
\sqrt{\frac{2}{3}} \ket{\downarrow} \nonumber
\\
\ket{\psi_3} & = & \sqrt{\frac{1}{3}} \ket{\uparrow} +
\sqrt{\frac{2}{3}} e^{2i\pi/3} \ket{\downarrow} \nonumber
\\
\ket{\psi_4} & = & \sqrt{\frac{1}{3}} \ket{\uparrow} +
\sqrt{\frac{2}{3}} e^{-2i\pi/3} \ket{\downarrow},
\end{eqnarray}
where $\ket{\overline{\psi}_i}$ represents the quantum state
orthogonal to $\ket{\psi_i}$, and $\epsilon$ quantifies the
strength of the measurement.

\begin{figure}[h]
\includegraphics[width=4.2cm]{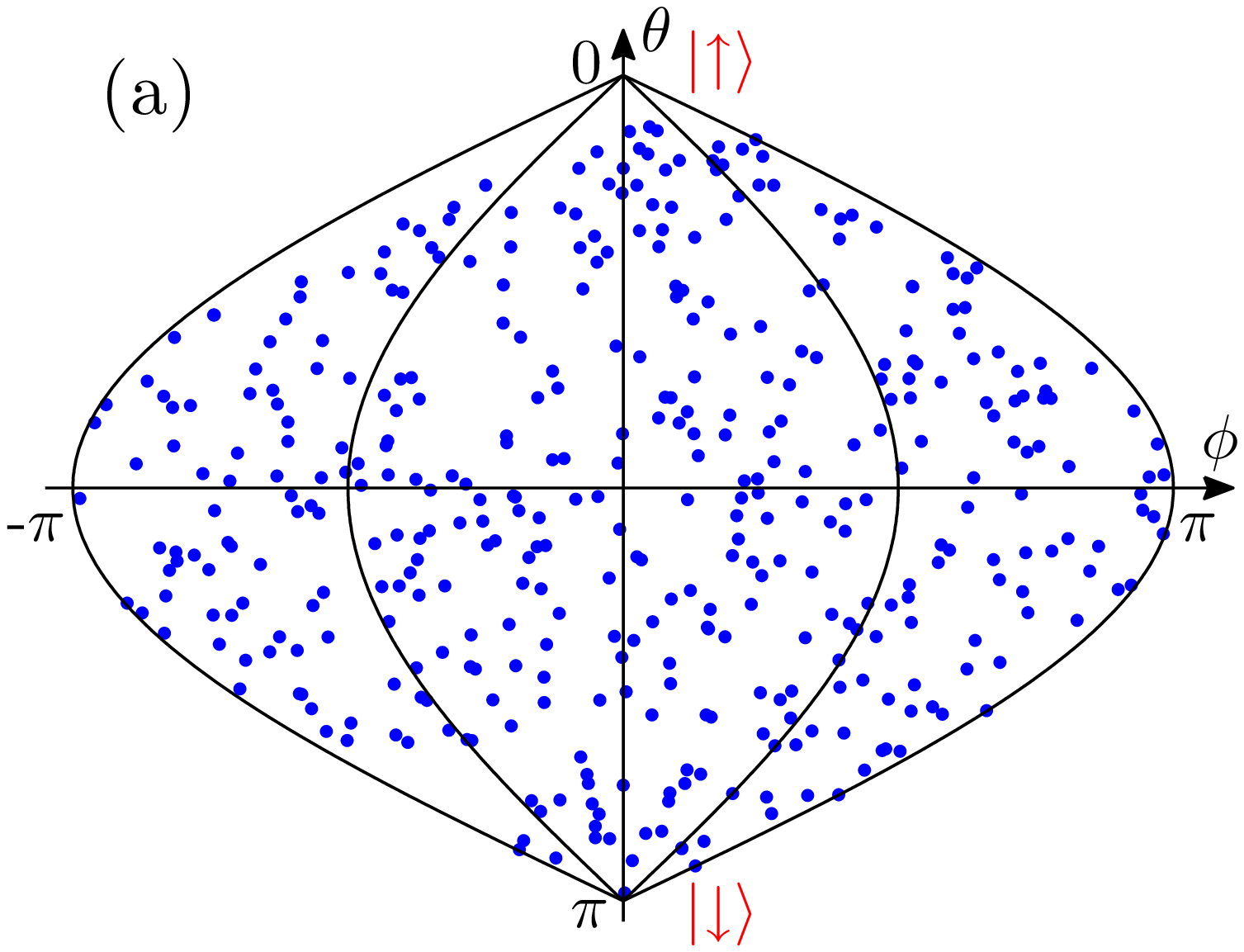}
\includegraphics[width=4.2cm]{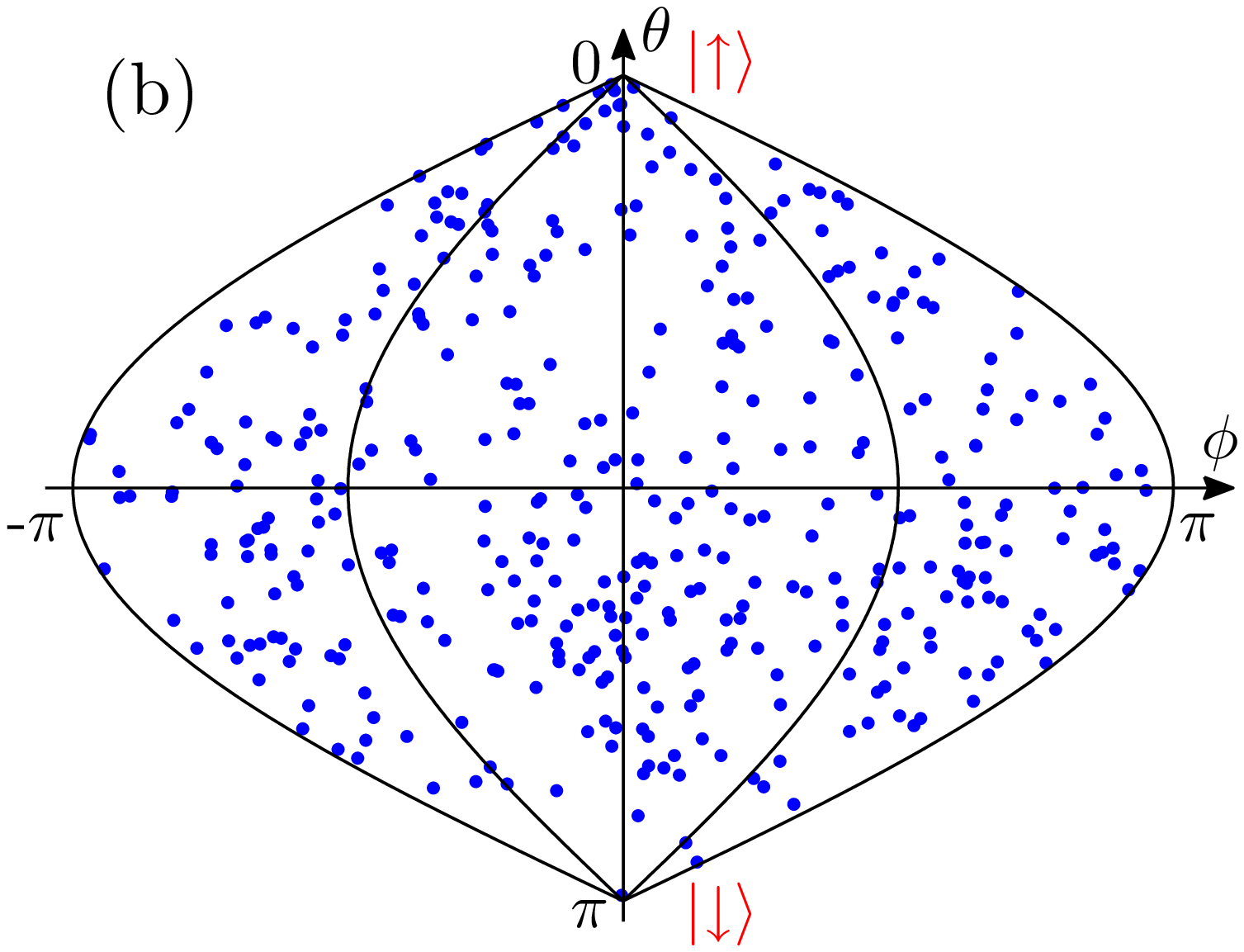}
\includegraphics[width=4.2cm]{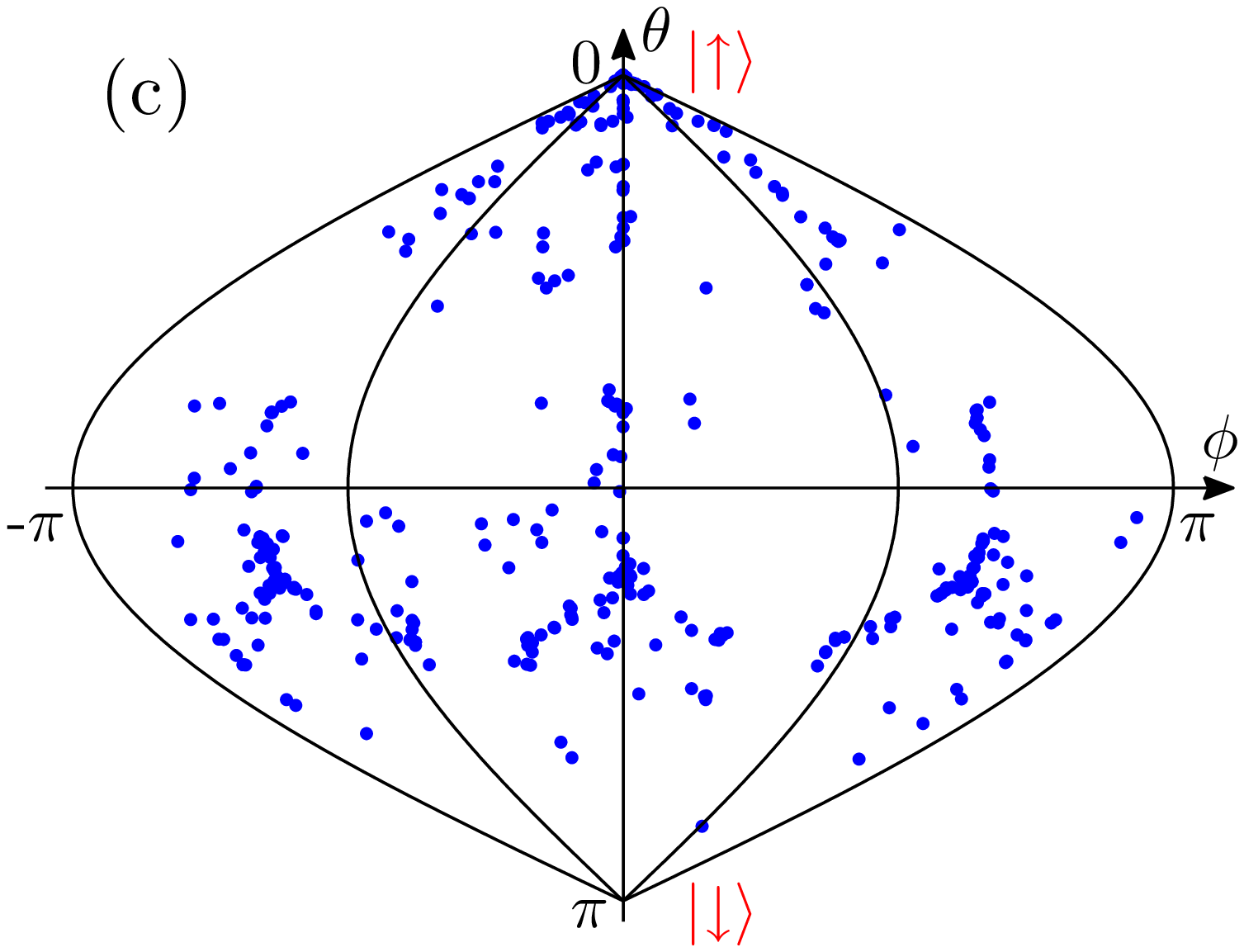}
\includegraphics[width=4.2cm]{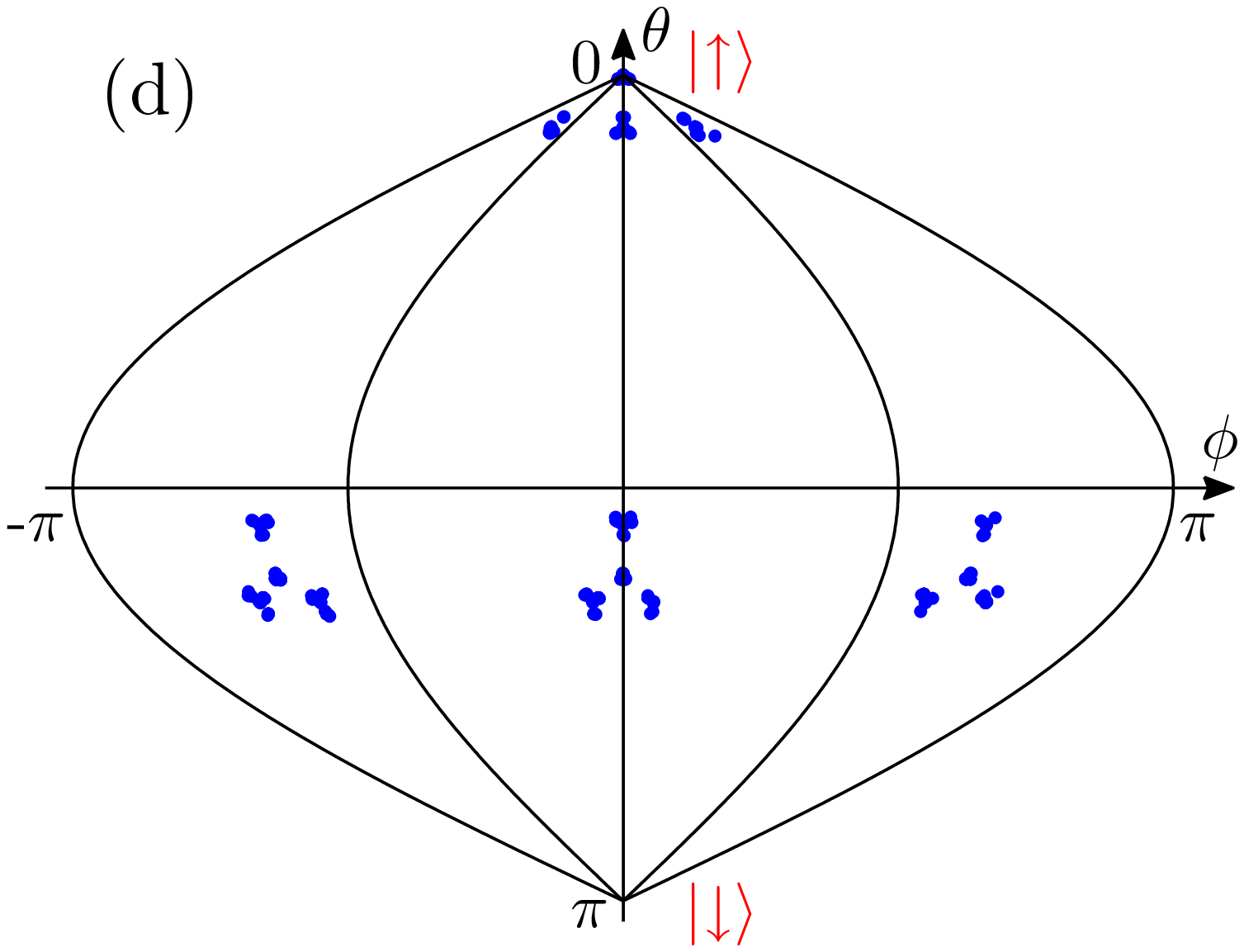}
\caption{(Color online) The distribution of states obtained from
repeated measurement using a SIC-POVM. The different panels
correspond to different measurement strengths: $\epsilon=0.1$ (a),
0.5 (b), 0.9 (c) and 0.99 (d). In each panel 400 point are
plotted. Each point represents the values of the spherical
coordinates $\theta$ and $\phi$ of the quantum state $\ket{\psi}$,
plotted using a map-like scheme designed not to distort the areas
of different regions on the surface of the Bloch sphere. In order
to obtain a distribution that is representative of the long-time
steady state, each simulation involves a sequence of $10^5$
measurement steps. The 400 points in each panel represent the
quantum state at equally spaced measurement steps, i.e.~we plot
the coordinates of the state after measurement step number 250,
500, 750, ... The results shown in this figure are independent of
the initial state used in the calculations.}
\label{Fig:SICPOVM}
\end{figure}

In Fig.~\ref{Fig:SICPOVM} we plot the coordinates of a quantum
state that is repeatedly measured using the SIC-POVM given above.
If one consider the case $\epsilon=1$ (not plotted), only the four
states $\ket{\psi_1}$, $\ket{\psi_2}$, $\ket{\psi_3}$,
$\ket{\psi_4}$ are realized as post-measurement states. It should
be noted here that (unlike von-Neumann measurements) even if the
state is one of these four states at a given step, repeated
measurement will not necessarily produce the same outcome; it is
possible that in the next step a different outcome is obtained,
and the state of the system changes to the corresponding state. As
can be seen clearly from Fig.~\ref{Fig:SICPOVM}(d), for values of
$\epsilon$ that are close to one, e.g.~at $\epsilon=0.99$, there
are 16 possible states that appear as a result of the repeated
measurements: the original four states (i.e.~the ones that appear
for $\epsilon=1$), and three more states surrounding each one of
the original four states. These four sets of three additional
states correspond to obtaining the outcome $i$ given that the
state was (approximately) given by $\ket{\psi_j}$ in the previous
step (with all the different combinations of $i,j=1,2,3,4$). When
$\epsilon$ is reduced to 0.9 [Fig.~\ref{Fig:SICPOVM}(c)], the
sixteen-point structure cannot be seen anymore, but we can clearly
identify that the points are concentrated in four regions on the
surface of the Bloch sphere. When $\epsilon$ is reduced to 0.5 or
below, we can see that the states now cover the entire Bloch
sphere, such that any state can be prepared
\cite{CommentRegardingFiniteEpsilon}. The reason why the states
are no longer concentrated around the four outcome states is that
for small values of $\epsilon$ the projection in each measurement
step is only partial, and the state undergoes a rather stochastic
motion on the Bloch sphere, allowing it to access all regions on
the surface of the Bloch sphere. It should be emphasized here that
even though the motion of the state is stochastic, one can keep
track of this motion given the record of the outcomes that have
been obtained in the sequence of measurements. One is therefore
able to tell if the state hits the target state $\ket{\psi_T}$ (up
to the accepted tolerance level). It is therefor possible to
prepare any target state $\ket{\psi_T}$ by repeatedly performing
measurements using a single SIC-POVM, provided that $\epsilon<1$.

One can obtain a quick estimate for the time required to prepare
an arbitrary target state by assuming that the state after each
measurement step is randomly located on the Bloch sphere,
independently of the measurement history. If the error tolerance
is set such that a deviation by an angle smaller than $\delta$ is
accepted, then one can divide the solid angle of the entire Bloch
sphere, i.e.~$4\pi$, by the solid angle defined by the deviation
$\delta$, i.e.~$\pi\delta^2$. The resulting estimate is that it
would take on average $4/\delta^2$ steps in order to prepare an
arbitrary target state.

\begin{figure}[h]
\includegraphics[width=5.0cm]{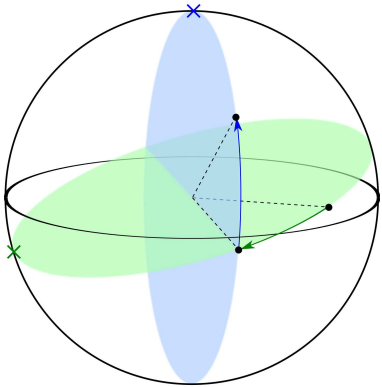}
\caption{(Color online) Schematic diagram showing how a given
initial state can be transformed into a given target state after
two applications of the SIC-POVM with appropriately chosen
measurement strengths. The first measurement brings the quantum
state into the plane defined by the target state and one of the
SIC-POVM states $\ket{\psi_i}$, and the second measurement results
in the target state. It should be emphasized that the success of
each one of these two steps is probabilistic.}
\label{Fig:SchematicTwoMeasurements}
\end{figure}

The above estimate for the average preparation time assumed weak
measurements. Alternatively one could assume a tunable measurement
strength in order to speed up the state-preparation process. We
have used numerical calculations to verify the following
statement: Given any initial state, it is possible to end up in
any given target state after at most three measurements, provided
that the measurement strengths are tuned properly and one obtains
the correct measurement outcomes. A typical example, where only
two steps are needed, is illustrated in
Fig.~\ref{Fig:SchematicTwoMeasurements}. One can therefore
conclude that it should be possible to prepare any desired target
state with high probability in under twenty measurement steps. We
shall not go into any detailed calculations of the average
preparation time here. It should also be noted in this context
that a SIC-POVM with a large value of $\epsilon$ cannot be
obtained by repeating a SIC-POVM with a smaller value of
$\epsilon$. Instead, performing two of these four-outcome POVMs
results in a sixteen-outcome POVM.

\section{Discussion and Conclusion}

In this paper we have given two examples demonstrating the
possibility of actively using the measurement-induced back-action
on a quantum system for purposes of manipulating the system, in
particular preparing an arbitrary quantum state.

In one case we have shown how measurements of the observables
$\sigma_x$, $\sigma_y$ and $\sigma_z$ can be used for
arbitrary-state preparation. Since the protocol is based on simple
principle of Euclidean geometry, it is not necessary that the
three measurement axes be orthogonal. The protocol can be
straightforwardly modified to work for any set of three linearly
independent axes. We have also shown that one does not need to
have three different measurements as available resources. Even a
single measurement, the SIC-POVM, was sufficient to prepare an
arbitrary quantum state.

The examples presented in this paper demonstrate that the only
requirement for the available measurements is that they must be
informationally complete, i.e.~they contain $d^2$ linearly
independent measurement operators (where $d$ is the size of the
Hilbert space). The above statement applies for weak measurements,
where the measurement only slightly modifies the quantum state. As
explained in Sec.~III, strong measurements can be more limited
than weak measurements in terms of the number of quantum states
that they can be used to prepare.

Our results also demonstrate that measurements can play an active
role in quantum control. This point is particularly important in
systems where it might be easier to perform measurements rather
than apply unitary operations. In such a case, one could design
the feedback-control protocol to rely more heavily or exclusively
on measurements.

In this context it should be noted that throughout this work we
have assumed minimally disturbing measurements, i.e.~measurements
whose measurement operators are the square-roots of the
corresponding POVM elements \cite{WisemanBook}. In other words,
the measurement does not induce any unitary operation on the
quantum state, except for the the quantum back-action associated
with the gain of information. The evolution of the quantum state
is therefore solely due to the gain of information in the
different measurement steps. The assumption of minimally
disturbing measurements also implies that we have neglected any
classical noise that adds some amount of uncertainty to the
post-measurement state, i.e.~even if the pre-measurement state is
pure the post-measurement state can generally be mixed. Any such
noise would reduce the fidelity of the prepared state.

In control theory, including quantum control theory, an important
question is the minimum number of available operations that are
needed in order to fully control the system. Much work has been
done on the minimum requirements for unitary operations required
to fully control a quantum system, and it is well known that any
two infinitesimally small, linearly independent rotations are
sufficient to generate any finite unitary operation on a single
qubit \cite{Burgarth}. In a similar spirit, it would be
interesting to understand the minimum requirements on measurements
that one needs in order to fully control a system, and what
requirements are needed for the case where one has a combination
of measurements and unitary operations. The above questions
concerning the minimum requirements for controllability can be
become mathematically challenging when dealing with quantum
systems with large Hilbert spaces, in contrast to the two-level
system considered in this work.

This work was supported in part by the National Security Agency
(NSA), the Army Research Office (ARO), the Laboratory for Physical
Sciences (LPS), the Defense Advanced Research Projects Agency
(DARPA), the Air Force Office of Scientific Research (AFOSR),
National Science Foundation (NSF) grant No.~0726909, JSPS-RFBR
contract No. 09-02-92114, Grant-in-Aid for Scientific Research
(S), MEXT Kakenhi on Quantum Cybernetics, and the Funding Program
for Innovative R\&D on Science and Technology (FIRST).


\begin{thebibliography}{99}

\bibitem{ControlSystemBook} See e.g. R. C. Dorf and R. H. Bishop,
{\it Modern Control Systems} (Prentice Hall, 2007).

\bibitem{WisemanBook} H. M. Wiseman and G. J. Milburn,
{\it Quantum Measurement and Control} (Cambridge University Press,
2009).

\bibitem{Brif} C. Brif, R. Chakrabarti, and H. Rabitz, New
J. Phys. {\bf 12}, 075008 (2010).

\bibitem{Peirce} It is also worth mentioning that open-loop feedback
control in quantum systems, where the problem of
measurement-induced backaction does not arise, has also been
studied extensively in the past; see e.g. A. P. Peirce, M. A.
Dahleh, and H. Rabitz, Phys. Rev. A {\bf 37}, 4950 (1988); W. S.
Warren, H. Rabitz, and M. Dahleh, Science {\bf 259}, 1581 (1993);
M. Shapiro and P. Brumer, {\it Principles of the Quantum Control
of Molecular Processes} (Wiley, Hoboken, 2003); D. D'Alessandro,
{\it Introduction to Quantum Control and Dynamics} (Chapman \&
Hall, Boca Raton, 2008).

\bibitem{Belavkin} V. P. Belavkin, Autom. Rem. Control {\bf 44}, 178
(1983).

\bibitem{Wiseman} H. M. Wiseman and G. J. Milburn, Phys. Rev. Lett.
{\bf 70}, 548 (1993).

\bibitem{Roa} L. Roa, A. Delgado, M. L. Ladr\'on de Guevara, and A. B.
Klimov, Phys. Rev. A {\bf 73}, 012322 (2006); L. Roa, M. L.
Ladr\'on de Guevara, A. Delgado, G. Olivares-Renter\'ia, and A. B.
Klimov, J. Phys. Conf. Series {\bf 84}, 012017 (2007).

\bibitem{Pechen} A. Pechen, N. Il'in, F. Shuang, and H. Rabitz, Phys. Rev.
A {\bf 74}, 052102 (2006).

\bibitem{Jacobs} K. Jacobs, New J. Phys. {\bf 12}, 043005 (2010).

\bibitem{Hida} For example, these ideas were used in a proposal
for the preparation of entangled states of two particles; Y. Hida,
H. Nakazato, K. Yuasa, and Y. Omar, Phys. Rev. A {\bf 80}, 012310
(2009); K. Yuasa, D. Burgarth, V. Giovannetti, and H. Nakazato,
New J. Phys. {\bf 11}, 123027 (2009).

\bibitem{vonNeumann} In fact, the basic idea of this procedure can
be found in J. von Neumann, {\it Mathematical Foundations of
Quantum Mechanics} (Princeton University Press, 1996).

\bibitem{Ashhab} See e.g. S. Ashhab, J. Q. You, and F. Nori,
Phys. Rev. A {\bf 79}, 032317 (2009); New J. Phys. {\bf 11},
083017 (2009); Phys. Scr. {\bf T137}, 014005 (2009).

\bibitem{CommentRegardingFiniteEpsilon} In fact, since there does not
seem to be any reason to have singular behaviour for any
intermediate value of $\epsilon$, we suspect that any state on the
Bloch sphere can be prepared using the SIC-POVM with any value of
$\epsilon$ that is smaller than one. However, for values of
$\epsilon$ that are very close to one the resulting state
distribution is highly concentrated around four points. In
Fig.~\ref{Fig:SICPOVM}, where only 400 points are shown in each
panel, the plotted points are all close to the four concentration
points. For example we have plotted a figure with the same
parameters as Fig.~\ref{Fig:SICPOVM}(c) but with $10^7$ points,
and in that case the points cover the entire Bloch sphere.

\bibitem{Burgarth} There has also been some recent work on
controlling a multi-qubit system by controlling a small subset of
the qubits; see e.g. D. Burgarth, K. Maruyama, M. Murphy, S.
Montangero, T. Calarco, F. Nori, M. B. Plenio, Phys. Rev. A {\bf
81}, 040303(R) (2010).

\end{thebibliography}
\end{document}